# Multiscale mechanical modeling of skeletal muscle: a systemic review of the literature

Aude Loumeaud[1,2], Philippe Pouletaut[2], Sabine F. Bensamoun[2],

Daniel George[1], Simon Chatelin[1*]

[1]*University of Strasbourg, CNRS, Inserm, ICube UMR 7357, Strasbourg, France*
[2]*Université de technologie de Compiègne, CNRS, BMBI UMR 7338, Compiègne, France*
*Corresponding author: schatelin@unistra.fr

**ORCID:**
Aude Loumeaud (0009-0005-7987-867X); Philippe Pouletaut (0000-0001-9700-040X); Sabine F. Bensamoun (0000-0003-4290-0173); Daniel George (0000-0003-3310-8141); Simon Chatelin (0000-0001-8999-3608)

**Abstract.** Purpose: From the myofibrils to the whole muscle scale, muscle micro-constituents exhibit passive and active mechanical properties, potentially coupled to electrical, chemical, and thermal properties. Experimental characterization of some of these properties is currently not available for all muscle constituents. Multiscale multiphysics models have recently gained interest as a numerical alternative to investigate the healthy and diseased physiological behavior of the skeletal muscle.

Methods: This paper refers to the multiscale mechanical models proposed in the literature to investigate the mechanical properties and behavior of skeletal muscles. More specifically, we focus on the scale transition methods, constitutive laws and experimental data implemented in these models.

Results: Using scale transition methods such as homogenization, coupled to appropriate constitutive behavior of the constituents, these models explore the mechanisms of ageing, myopathies, sportive injuries, and muscle contraction.

Conclusion: Emerging trends include the development of multiphysics simulations and the coupling of modeling with the acquisition of experimental data at different scales, with increasing focus to little known constituents such as the extracellular matrix and the protein titin.

**Keywords.** Multiscale, Multiphysics, Numerical modeling, Skeletal muscle, Biomechanics, Homogenization

**Statements and Declarations**

**Fundings.** The work was supported by the Research Department of UTC within the framework of AMI Internationaln and also by the Interdisciplinary Thematic Institute ITI HealthTech, as part of the ITI 2021–2028 program of the University of Strasbourg, CNRS and Inserm, through IdEx Unistra (Grant number ANR-10-IDEX-0002) and SFRI (STRAT'US project, Grant number ANR-20-SFRI-0012).

**Competing Interests.** The authors have no relevant financial or non-financial interests to disclose.

**Author Contributions**. All authors contributed to the realization of this review. The first and last drafts of the manuscript were written by Aude Loumeaud and all authors commented on the successive versions of the manuscript. All authors have read and approved the final manuscript.

**Ethics Approval.** The authors confirms that the study was approved by the appropriate institutional and national research ethics committee and certify that the study was performed in accordance with the ethical standards as laid down in the 1964 Declaration of Helsinki and its later amendments or comparable ethical standards.

**Consent to Participate.** The authors confirm that informed consent to participate in the study has been obtained from participants when necessary.

**Consent to Publish.** Individuals consent to participate in this study, and do not object to having their data published in a journal article. The authors attest to have sought consent from individuals to publish their data prior to submitting this paper.

**Data Availability Statement.** The authors confirm that prior consent has been obtained for all data as soon as necessary.

# 1. Introduction

Skeletal muscles present a multiscale architecture [1,2], as illustrated in Fig 1. To each scale corresponds a specific architecture with associated constituents, each of which is likely to contribute to the active or passive mechanical properties of the overall skeletal muscle [3,4]. At the macroscopic scale (of the order of centimeter), the muscle is constituted of several adjacent mesoscale fascicles. These fascicles in turn contain groups of tightly packed muscle fibers or muscle cells that constitute the microscale (from 10 to 100 µm in diameter). Muscle fibers can be arranged in different architectures within the muscle depending on the muscle's function in the body [5]. All the aforementioned structures are separated by thin layers of connective tissue known as extracellular matrix [6,7]. Fibers are constituted by the submicron scale myofibrils (from 1 to 2 µm in diameter) and cellular components. Each of these myofibrils is an assembly of sarcomeres in series that contain proteins such as actin, titin and myosin. Other cellular components include the sarcoplasmic reticulum, that propagates neural signal as a chemical potential down to the sarcomeres. They are physically linked to extracellular structures through costameres [8–10]. On the other end of this linkage, the extracellular matrix consists in collagen and elastin fibers embedded in a matrix of proteoglycans [11,12].

The mechanical properties of the smaller scale components result in a global anisotropic viscohyperelastic behavior of the skeletal muscle [7,1,13]. The passive mechanical properties in the skeletal muscle are attributed to two proteins, namely titin in the fibers [14–16] and collagen in the extracellular matrix [1,17–20]. Transmembrane proteins such as dystrophin that belong to the costameres also contribute to lateral force transmission [1,21] between muscle fibers and extracellular matrix. During muscle activation, the change in mechanical properties is due to the sarcomere contraction attributed to the binding of proteins actin and myosin, namely the cross-bridge cycling theory. Some evidence also suggests an active bound formation between titin and actin during activation, called force enhancement [22]. The skeletal muscle mechanical properties are related to the muscle's physiological state. To study the functional or the structural properties of the skeletal muscle, *in vivo*, *ex vivo* or *in vitro*, experiments can be conducted. These investigations are however limited due to technical difficulties when it comes to measure the mechanical properties at smaller scales, especially in human patients [23]. Thus, tools such as multiscale biomechanical modeling are implemented [24,25]. These tools first establish the mechanical link over scales in the organ using several approaches such as analytical homogenization, numerical homogenization, and averaged results, which are accounting for the different constituents of the microstructure and their specific geometry at each scale. Then, the mechanical constitutive laws for the constituents must be identified, including passive and active components as well as potential multiphysics components (such as electrical, chemical or thermal specificities).

In this study, we describe the current state of the art in multiscale and multiphysics biomechanical numerical modeling, with a focus on models describing the skeletal muscle mechanical behavior from myofibrils to the whole organ, excluding smaller than submicron [26] and larger than macroscopic [27–29] scales. After describing the main modeling approaches, we will expose and compare the mechanical constitutive laws implemented at the different scales for passive and active mechanical models as well as multiphysics models.

# 2. Multiscale modeling

To establish the link from one scale to another, several approaches have been proposed in the literature for skeletal muscle modeling, which can be classified into three main categories: analytical homogenization, numerical homogenization, and averaged results.

## 2.1. Model geometry

Multiscale modeling implies a definition of a simplified geometry that is representative of the skeletal muscle at the considered scale, based on hypotheses on the anatomical constituents [30]. Here we discuss implementation of realistic and idealized geometries. Realistic geometries tend to reproduce medical imaging acquisitions [31] and are implemented in skeletal muscle multiscale models mostly from microscopy cross section images [32–37]. Kuravi and colleagues propose a 3D realistic geometry obtained from registration and segmentation protocols on successive histological cross sections, resulting in a 3D reconstruction of a cubic muscle sample at the microscale [32,33]. Realistic geometries can also be AI generated [38].

However, most multiscale models do not use microscopy images directly, but rather simplified geometries either in 1D, 2D or 3D (Fig. 2). 1D elements are used to represent constituents at the submicron scale such as titin represented with springs [39–41], or myofibrils and collagen represented with bars [42] (Fig2.A). 1D muscle fibers are modeled at the microscale as springs [43–45] and as bars [39–41] (Fig2.B). At the microscopic scale, some authors assume that muscle fibers are parallel to each other. Thus 2D plane geometries representing a muscle cross section are often extruded alongside the direction orthogonal to the plane, generating a 3D geometry with regular arrangements of parallel muscle fibers such as perfectly circular [46–48] (Fig 2.C) or hexagonal [49,50] (Fig 2.D). Perfectly circular arrangements are simple of use especially considering analytical homogenization methods, as these methods often consider elliptical or cylindrical fiber reinforced composites [51]. However, the maximum fiber packing or fiber volume fraction (FVF) that can be obtained using this arrangement is 90.7% whereas the muscle fiber FVF in healthy muscles reaches 95%. Moreover, this geometry creates local section variations in the ECM, which are not representative of the real microstructure [52]. Hexagonal muscle fibers overcome these drawbacks, yet they consider fibers with equal geometries and areas, while the real microstructure is more random and contains fibers with different sizes [53]. Thus, an increasing trend is the use of randomized regular arrangements such as Voronoi tessellations [54]. Voronoi tessellations or a Voronoi diagram constitute in partitions of a plane into convex regions also called cells. The cells are defined by a given set of points called Voronoi seeds or centroids. Each Voronoi region is defined by a seed. Any point belonging to that region is closer to that seed than to any other seed present in the set. The obtained regions vary in size and shapes. Periodic Voronoi tessellations are obtained by duplicating the seeds contained in a 1x1 region into a 3x3 grid and intersecting the obtained Voronoi diagram by a 1x1 bounding box (Fig 2.E). This specific arrangement has been used for periodic numerical homogenization [55–61]. Some authors also define curved-edge Voronoi tessellations [32,33,57] using different methods to round up the obtained Voronoi regions, arguing that the realistic geometry does not contain sharp angles. At the mesoscale, fascicles and extracellular matrix have been also represented using Voronoi tessellations [62]. At the macroscale, idealized 3D muscle geometries include rectangular parallelepiped and idealized fusiform muscle shape [43–45,55] (Fig 2.F).

In the specific context of homogenization, the representative geometry is named unit cell or Representative Volume Element (RVE) [30,63]. Recently, tools that can be used to generate RVE in Finite Element software have been developed [62,64,65]. They allow to automatically generate microscale circular arrangements [65], submicron scale to mesoscale arrangements including Voronoi tessellations [62] and periodic boundary conditions [64], that have specific application for numerical homogenization or averaged values scale transitions.

### 2.2. Scale transition approaches

Homogenization is currently the most widely used approach for multiscale modeling of skeletal muscle. Homogenization methods have been developed for the study of multiscale materials [66–70]. Thus, these methods consider two separate spatial scales: the microscopic scale where the material presents a heterogenous microstructure, and the macroscopic scale, where the material can be supposed homogenous, as the heterogeneities' dimensions are considered neglectable. Homogenization methods approximate the mechanical behavior of the macroscopic scale material by computing the effective mechanical properties of an equivalent homogeneous material which are representative of the microscopic constituents' behavior. They provide a link from the microscopic scale to the macroscopic scale (homogenization) and sometimes an inverse link from the macroscopic scale to the microscopic scale (localization). There exists two main homogenization methods families, namely the numerical homogenization methods [71], which require the use of specific simulation tools, and the analytical homogenization methods [72] which can be applied at the different scales at which we seek to model skeletal muscle.

Homogenization methods require assumptions and definition of i) the heterogeneous material's geometry at the microscopic scale (RVE), ii) the constituent's constitutive behaviors, and iii) the interactions between the constituents at the microscopic scale. From the submicron scale to the mesoscale, skeletal muscle can be modeled as a fiber reinforced composite, with a FVF ranging from 90 to 95% for healthy muscles at the microscale. The muscle's microconstituents exhibit an anisotropic behavior [73–75] but are often assumed transversely isotropic in multiscale models. The aforementioned features constitute a current limitation for the use of most analytical homogenization methods developed for the investigation of fiber reinforced composites. Amongst these methods, the Mean Field Homogenization (MFH) methods [51] are based on the so-called dilute dispersion assumption, which considers that the fibers should be contained in dilute quantities in a very large matrix medium, therefore excluding materials with high FVF. Thus, other homogenization methods have been applied to skeletal muscle modeling. The Voigt and Reuss methods, also named rule of mixture laws, provide respectively upper and lower bounds for the behavior of the equivalent homogenous material in the case of a composite reinforced with long parallel fibers [69,70]. The Voigt method [70] assumes an isometric strain in the composite and is consequently used for load cases in the direction of the fibers. Its counterpart is the Reuss method [69] which considers isometric stress in the composite. The advantage of these methods is their simplicity, as they are based only on the FVF, however this implies that they do not take the microscopic geometry into account. Due to this simplicity, homogenized skeletal muscle models often use the Voigt homogenization [46,50,59,76]. Another method based on asymptotic homogenization [77,78] has been used in skeletal muscle modeling [42]. Briefly, asymptotic homogenization assumes that at least one direction of space at the microscopic scale is periodic. The method relies on the displacement field's description at the microscopic scale using fast and slow space variables to account for this periodicity, and then on the application of finite series development to obtain an equation system. [77] developed an estimate for graphite sheets, that corresponds to the linearized behavior of the composite around a pre-stressed configuration. This method provides a better estimate than the Voigt and Reuss methods, at the cost of increased complexity. One last family of homogenization methods used in the context of skeletal muscle multiscale modeling is the second order methods, that have been developed for the modeling of nonlinear elastic composites. Second order methods based on estimates for sequentially laminated composites [79,80] have been implemented [55] and recently developed [81] for biological materials such as skeletal muscles. Through these methods, high FVF composite including constituents with transversely isotropic behavior can be replicated. [81] developed an estimate considering an anisotropic behavior for all microscopic constituents of a fiber reinforced composite. However, these methods exhibit great complexity and cannot account for complex geometries at the microscale.

Numerical or computational homogenization methods help overcome the limitations associated with analytical approaches to predicting the behavior of heterogeneous materials with complex geometries and solicitations. These homogenization methods are based on simulations of a discretized geometry. They can be separated in two categories, namely hierarchical and concurrent [71]. Hierarchical methods consider different simulations for the different scales, and information is passed from one simulation to another. Usually, one macroscopic scale simulation is linked to several microscopic scale simulations. The hierarchical approach is implemented through the use of Finite Element Modeling for skeletal muscle homogenization [36,48,50,55–59], mostly through the use of computational periodic homogenization [82] which requires the discretized microscopic scale geometry to be periodic in all space directions. Conversely, concurrent methods consider different scales simultaneously in different subvolumes of the same simulation and these scales may exchange information in hand-shake regions. Some authors also developed a concurrent numerical homogenization method using two different meshes in the same simulation and computing variable interpolation between the two structures [83] for muscle simulation [39–41,84]. Hierarchical methods can be used to increase the model resolution in the case of damaged areas. However, they do not allow to model Fracture Process Zones (FPZ) to simulate muscle tear for example. This is made possible by concurrent approaches that render a more precise interaction between the scales. Both methods are computationally expensive.

Some multiscale models do not use homogenization methods. Modeling approaches such as Hill's model [34,35,85,86] draw the link between scales as it separates the mechanical contribution of each microscale constituent. Another approach, namely micromechanical modeling, considers large microscale geometries representative of the skeletal muscle and considers the behavior of these composites identical to the macroscale muscle's behavior. This information on the macroscale can be obtained through the average value of stresses and strains over the considered representative volume or surface. This approach is simple and allows a more complex representation of geometry and behavior. This method is used to obtain information on the microstructure, however phenomena happening at the macroscale (i.e. FPZ) cannot be modeled with this approach, as it doesn't provide localization.

## 3. Constitutive laws for the passive behavior

Some of the multiscale mechanical models of skeletal muscle include multiphysics parts, such as chemical [86,39–41,44,45,87] or electrical ones [86,39–41,44,45], to mimic the muscle activation in an active component. It is therefore essential to distinguish, at each scale of skeletal muscle modeling, between passive and active components, both of which will influence the mechanical behavior of the muscle at higher scales. In this section, the passive mechanical constitutive laws implemented at the different scales are first described, as to draw a scale-wise comparison in a second part. For further details, please refer to Appendix A for selective bibliography.

## 3.1. Passive mechanical constitutive laws at lower scales

At the submicron scale, constitutive laws are developed for myofibrils or more specifically titin. Titin is modeled as a linear spring [39–41] or a worm-like-chain model [88]. Myofibrils and collagen fibers are modeled as crimped fibrils in an isotropic matrix [42].

On the scale above, referred to here as microscale, muscle fibers constitutive laws depend on the modeling framework used. In Finite Difference Modeling (FDM), linear springs are used to represent muscle fibers in traction [43–45] and in compression [44,45]. Similarly, using Hill's model, the fiber is represented by 3 elements, including in series contractile and elastic elements , and a parallel elastic element, using exponential and quadratic hyperelastic formulations [34,52]. Using Finite Element Modeling (FEM), Reproducing Kernel Particle Method (RKPM) [37], or analytical frameworks, muscle fibers are usually modeled by separate isotropic and anisotropic contributions. The isotropic behavior accounts for cellular components other than myofibrils and is usually represented by simple hyperelastic laws such as Neo-Hookean [76,50,55,56,47,48,60,87], second order Mooney Rivlin [58], Yeoh model [46,49], or a quadratic polynomial function [37].

The muscle fiber anisotropic behavior is attributed to myofibrils and especially titin proteins. Constitutive laws are expressed as a function of the fiber stretch with a linear [50,55], quasi linear [42], exponential [37,58] or polynomial [46,49,56] shape, which coefficients are identified on muscle fiber uniaxial tension experimental data [4]. The fiber stretch is directly related to the Cauchy-Green tensor's invariant $I_4$, thus most models make use implicitly of invariant $I_4$ to model anisotropy. Using a more phenomenological approach, this invariant is explicitly used by [32,33] following the formulation proposed in [89]. In the same way, [57] use invariant $I_4$ in combination with invariant $I_5$. The use of invariant $I_5$ is justified by the need to represent shear deformations more accurately [90]. The Criscione invariants [91] have been used to represent shear deformations [36,38,59,92,93], as these invariants have a better physiological interpretability than the classical ones thus leading to more interpretable modeling results. The aforementioned approaches for muscle fibers constitutive modeling are summed up in Fig. 3. Some authors introduce corrective coefficients to take age into account [37,58]. The titin contribution in the fiber can also be found as a homogenized version of a submicron scale model [39–41]. To study rupture in fibers, a model of irreversible deformations and microcracks is proposed in [48].

ECM is represented as a linear elastic material [43–45], an isotropic material [35,46–48,60] or as a collagen fiber reinforced material [32,33,50,55,56,58,76]. Isotropic depictions of ECM include Yeoh [46] and first order Ogden [47,48,60] hyperelastic models. The Hill model [94] has also been extended to separate the muscle fibers and ECM contribution by adding a parallel stiffness to the previous 3D model [35] representing ECM by means of an exponential formulation.

As a fiber reinforced material, the ECM is modeled as an isotropic matrix reinforced with collagen fibers. As the collagen fibers are wavy and present a specific orientation in the tissue, formulations accounting for the waviness and uncrimping of the collagen fibers in soft tissues have been developed [88,95,96]. The isotropic matrix has been modeled considering a Neo-Hookean behavior [76,50,55,56] or a Mooney Rivlin model [58]. Authors systematically use collagen fiber orientations of either ±55° or ±59° relative to the muscle fibers' orientation. Two main families of methods have been described and compared in the literature to model the collagen fibers: the angular integration (AI) approach [97] and the generalized structure tensor (GST) approach [98]. The AI approach considers the mechanical behavior of each collagen fiber and then integrates this behavior over a volume where specific fiber orientation distributions can be defined. On the contrary, the GST approach uses a tensor called generalized structure tensor as to represent a fiber family's global orientation, and the mechanical behavior is defined per fiber family. This topic has been reviewed in [99]. The angular integration approach for skeletal muscles extracellular matrix is presented in [88]. More models use the AI approach [50,55,56,58,59,76] than the GST approach [32,33].

Recent ECM models considering collagen cross-linking in the ECM have been developed [100,101] as this feature directly influences the mechanical properties of ECM [1,18]. Moreover, as costameres connect the interior of the muscle fiber to the ECM, this leads to a phenomenon called lateral force transmission, where the fiber's force is transmitted through shear to the ECM. The ECM network then locally redistributes the loads into the fascicle through endomysium shear [1,102]. Consequently, shear modeling in the ECM has gained increasing interest in multiscale models. The previous approaches do not consider shear contribution of the collagen fibers into the extracellular matrix. Recent models aimed at including shear stress related components in ECM constitutive laws. [33] developed a specific skeletal muscle extracellular matrix behavior law based on the GST approach that also includes a volumetric contribution to penalize volume changes and a structural contribution to model shear. Similarly, [57] used a passive anisotropic hyperelastic contribution for extracellular matrix that includes the Cauchy-Green tensor's invariants I5 and I7, that was developed to better model shear in arteries [103].

Fat inclusions at the microscale are modeled as an isotropic neo-Hookean material [38,56]. Transmembrane proteins have also been included at the microscale [38] as nonlinear springs. The rupture in ECM, muscle fibers, and tendon fibers [104] can also be implemented as in [45].

Very few authors include fascicles as a realistic boundary material in multiscale models. To the best of our knowledge, no specific constitutive laws have been developed for the fascicles in the context of multiscale modeling. A general approach is the use of a rule of mixtures applied to micro constituent behaviors to determine the parameters of a hyperelastic constitutive law [35,38]. The parameters of a 3D Hill model have been identified by the response of a motor unit [86]. They differ from the fascicles as muscle fibers contained in the same motor unit do not necessarily belong to the same fascicle. The skeletal muscle has been represented with motor units grouped in parallel.

### 3.2. Validation on experimental data at the whole muscle scale

An important step in biomechanical modeling is the validation of numerical models on experimental data. For skeletal muscle multiscale modeling, a few datasets including mainly uniaxial tension and shear tests are used in this way [75,105,106] by comparison to their numerical replication [33,38,46,107], as detailed for the most frequently used experimental studies in the Table 1.

A specific difficulty in validating the models in both shear and compression loadings can be denoted. Implementation of shear loading allows to investigate conditions such as Duchenne Muscular Dystrophy [38,59], cerebral palsy [46,107] and ageing [37,58] that have an impact on the ECM. Compression has been modeled to study deep ulcer injuries in a model of hypoxia including the capillaries [87].

## 4. Constitutive laws for the active behavior

### 4.1. Active mechanical behavior

To get insights on neural stimulation and force generation within the skeletal muscle, an active contribution to the constitutive laws is added. Force generation in the skeletal muscle is attributed to proteins actin and myosin binding in the presence of calcium, namely the cross-bridge theory [110,111]. Experimental evidence also suggests an additional active force generated by formation of a bound between titin and actin during activation [13,22], that would be responsible for the residual force enhancement phenomenon first observed and described in [112]. To account for the bound formed between titin and actin in the presence of calcium, namely titin force enhancement, titin is modeled using the so-called "sticky spring mechanism" [39–41,113] that assumes that a certain region from the titin molecule named PEVK region binds to actin in the presence of calcium. In this model the titin is modeled as a linear spring parallel to the muscle fiber. Its free length gets reduced when titin binds to actin and its orientation changes due to the location of the bound, thus inducing additional efforts both in the fiber direction and in the cross-fiber directions.

Several approaches have been proposed to implement active behavior at the microscale. The active behavior is usually an additional component in the muscle fiber constitutive behavior generating specific stress or pressure along the muscle fiber axis [52]. A value representative of the active stress generated by a single fiber is affected, e.g. the maximal isometric force [32,33,49]. This additional active stress can also be modulated by the product of activation parameters, activation functions, normalized force-length, and force-velocity relationships. Activation can be represented by a single parameter α which values can range from 0 (non-activated) to 1 (full activation, all actin-myosin bounds are supposedly formed) [48,50,55,60]. Usually, activation functions are used, as to solve numerical issues [44] or to better represent the physiological phenomena leading to activation [39]. Activation functions can exhibit linear [46], trigonometric [44], or exponential [34,58] formulations as a function of time with possible implication of electrical and chemical phenomena. Heidlauf and colleagues [39–41] base their work on the excitation contraction coupling model of Shorten et al., (2007) [114]. Activation can also be spatially dependent, with specific fiber recruitment patterns [49] to mimic neural signals.

Force length relationships describe the experimentally observed relationship between strain along the fiber axis and active stress, as the force generated by sarcomeres depends on their initial elongation at the activation. They exhibit piecewise polynomial [39–41,44,47,48,58,60], trigonometric [46], exponential [34] formulations, or can be found expressed as a Weibull distribution [50,55]. These functions are usually normalized as to modulate the maximal isometric force in the muscle fiber, similarly to activation functions.

Some authors implement a force velocity relationship, which captures the strain rate dependence of the active behavior, thus the viscous effects, by a hyperbolic relation [94,34,50,55].

Titin force enhancement can also be considered at the microscale through an additional stress component [48]. To study the recruitment of motor units in isometric conditions, a model has been proposed based on the Huxley's formulation [86,115].

Considering the whole muscle, three kinds of muscular contractions have to be distinguished. Isometric contractions happen at fixed muscle length; concentric contraction leads the muscle to shorten as it produces force, while in eccentric contractions the muscle elongates due to a greater opposite force. As isometric contractions do not generate changes in the geometry, they are modeled in a different way than concentric and eccentric contractions. Most authors model isometric contractions. This load reproduces voluntary contractions in human patients [86]. Isometric contractions are also modeled to investigate phenomena such as trigger points i.e. a local knot of permanently contracted sarcomeres without any neural stimulation [60], titin force enhancement [39–41], and effect of activation patterns within a muscle fiber bundle [49]. Eccentric contractions are modeled in the context of injuries due to eccentric contractions [44,48]. To numerically implement eccentric contractions, authors perform stepwise uniaxial extensions followed by isometric contractions. Concentric contractions are modeled to reproduce an isovelocity shortening protocol for parameter identification purposes [34]. A case study of both concentric and eccentric contractions has been reported [47].

### 4.2. Inclusion of electrical, chemical and thermal couplings

Active multiscale mechanical muscle models often include neural signaling, thus implementing other physics such a chemistry or electricity.

Based on an excitation contraction coupling model [114], a first model proposes to describe the ion dynamics of the excitation contraction coupling, such as calcium, chloride, potassium and magnesium, as well as adenosine triphosphate (ATP) dynamics [39–41]. The ions channels carry currents that represent the propagation of the action potential through the muscle fiber, represented by bidomain equations.

Another model of motor unit recruitment representing voluntary and evoked recruitment of motor units has also been developed [86] based on [116]. The action potential generated in each motor unit is then linked to calcium ions release in the myoplasm.

These electrical contributions are linked to chemical contributions representing ion dynamics. Upon release from the sarcoplasmic reticulum, calcium ions bind to molecule troponin, forming cross-bridges, which are responsible for the sarcomere contraction. The concentrations of attached cross-bridges in the different states of the contraction are calculated and normalized by the maximum concentration of possible cross-bridges, which is associated to the maximal force [39–41,83]. Similarly, voluntary motoneuron recruitment and sarcoplasmic reticulum calcium release have been modeled in motor units [86], inducing 3 states in the motor unit, depending on the calcium concentration: activated, in relaxation or relaxed. Chemical coupling has also been used to model oxygen diffusion in muscle capillaries and hypoxia [87].

A thermomechanical model accounting for thermal treatments in trigger points in the skeletal muscle has been developed [60]. Each component (namely muscle fibers and ECM) is described by means of mechanical, thermomechanical, and conductive thermal constitutive contributions. Thermomechanical coupling terms account for the thermal dilatation of the materials. During muscle activation, the cross-bridges undergo endothermal cycles, and are thus considered as a negative heat source.

Multiphysics multiscale models are often implemented to model neural stimulation and action potential propagation within the muscle. Through separation of muscle fibers in motor unit [84], the electrical input is transformed into transmembrane ionic potentials and finally transmitted to the contractile elements through calcium release from the sarcoplasmic reticulum. Other original approaches include the study of hypoxia in fibers, modeling the oxygen diffusion from capillaries and mitochondrial metabolism [87]. A promising application is the thermomechanical study of muscles, giving insights on thermal therapies for muscles such as cryotherapy [60].

## 5. Conclusions

In the literature, there exist only a few multiphysics multiscale models. Firstly, electrical and chemical couplings require the use of a multiphysics computational framework. Multiscale skeletal muscle models mostly use the Finite Element software ABAQUS® SIMULIA (Dassault Systèmes SE, Vélizy-Villacoublay, France) and the Discrete Element software GranOO [117] even if some opensource alternatives have been developed, such as the library OpenCMISS [84,118–120]. Multiphysics simulations are very demanding in resources and computation time. Consequently, parallel computation and optimization strategies need to be implemented in the multiphysics frameworks [119,120]. These considerations constitute another complexity layer to the development of a multiphysics multiscale muscle model, thus constituting a current limitation in muscle multiscale modeling. Other multiphysics frameworks have been used for multiscale skeletal muscle models implementation such as COMSOL Multiphysics (COMSOL Inc., Stockholm, Sweden) [49] and FeBio (Musculoskeletal Research Laboratories, University of Utah, Salt Lake City, UT, USA) [121].

In an attempt to compare these models, we have analyzed the equivalent shear moduli in small deformations proposed by each of the models for their passive behaviors, whether for ECM or muscle fibers (Fig. 4). We show in green the regions where muscle fibers are stiffer than ECM for passive mechanical properties, and in orange the areas where this trend is reversed. It can be seen that, even if there are wide disparities in values (notably linked to the fact that active behavior or other chemical or thermal properties that are added are capable of modifying these values for a certain number of models), two major trends emerge: models using stiffnesses of the same order of magnitude between ECM and muscle fibers, and models with ECM 3 to 5 times stiffer than fibers in the context of passive behavior alone.

For identification and validation purposes, data from different species and studies is used in the same model. Very few authors have indeed acquired data at different scales on the same samples [3,122,123]. Moreover, passive mechanical tests at several scales are performed *ex vivo*, thus leading to uncertainties due to the samples' conservation and preparation. Furthermore, datasets across the literature are not performed at the same velocities, thus the viscoelastic effect [124,125] may become important when combining several studies for parameter identification [23]. As a result, the literature contains scarce experimental data suited for multiscale modeling. Consequently, a current trend is to acquire the experimental data needed for the identification and the validation of the model parameters [32,33,57,59,126]. The acquired datasets are extracted from *ex vivo* passive experiments. We recommend exploring *in vivo* imaging techniques that provide insight on mechanical parameters within the muscles, such as elastography [127–129]. Model of small animals face an additional experimental challenge due to the size of the animal [130], that leads to scarcer datasets across the literature. Across the current models, the combination of data from different species can be observed, although differences can exist between them in anatomical features [5,131]. Some efforts have been made to compare the anatomy of skeletal muscles across species [131]; however, to our knowledge, mechanical properties have yet to be compared in the same way.

# Appendix A Main multiscale muscle models

The main multiscale skeletal muscle models reviewed here are summarized in Table 2.

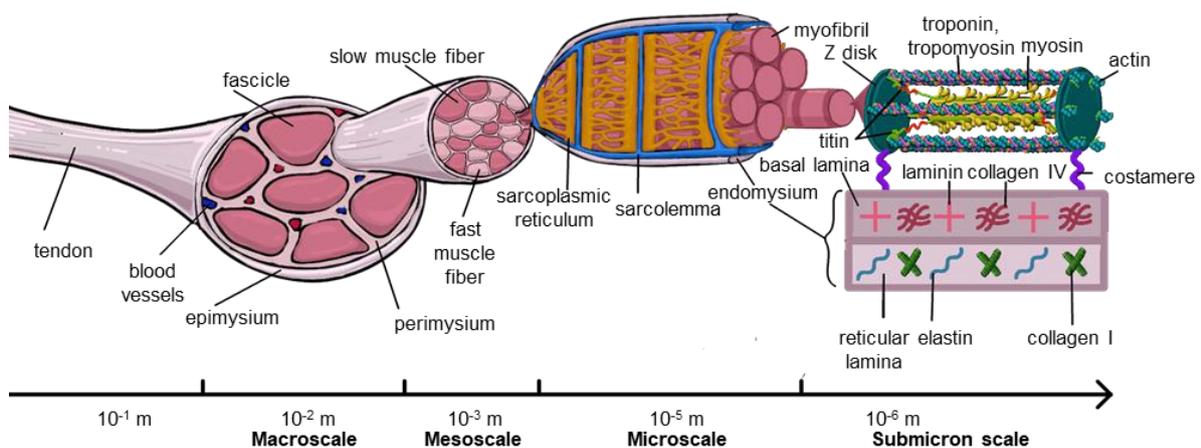

**Fig.1** Schematic representation of the multiscale architecture of the striated skeletal muscle [descriptive caption: a scheme representing the muscle and epimysium at the macroscale. The muscle constitutes of fascicles, blood vessels and perimysium. At the mesoscale, a fascicle is represented. It is surrounded by perimysium and it contains fast and slow fibers, as well as endomysium. At the microscale, a fiber constituted of myofibrils surrounded by the sarcoplasmic reticulum, the sarcolemma, and endomysium is shown. At the submicron scale a sarcomere from the myofibril, its link to the endomysium, and the endomysium are shown. More specifically, several proteins actin, titin, myosin, troponin, tropomyosin, costameres, laminin, collagen IV, elastin and collagen I are represented.]

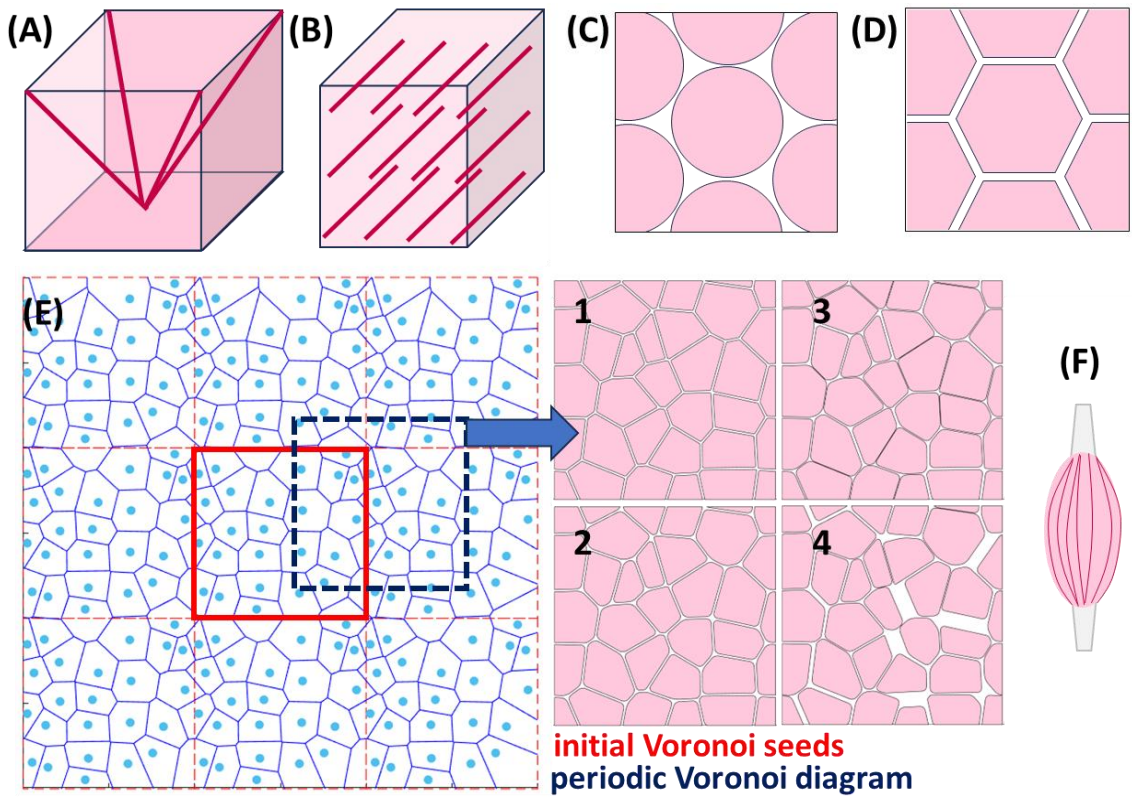

**Fig. 2** Idealized geometries to represent skeletal muscle: (A) myofibrils represented as trusses in a 3D geometry, (B) parallel 1D fibers embedded in a 3D muscle geometry (C) regular circular arrangement, (D) regular hexagonal arrangement, (E) periodic Voronoi diagram with constant ECM width (1), rounded up diagram (2) and variable ECM width (3 and 4), (F) fusiform muscle geometry [descriptive caption: this figure shows different simplified geometries used to describe the skeletal muscle in 2D and in 3D]

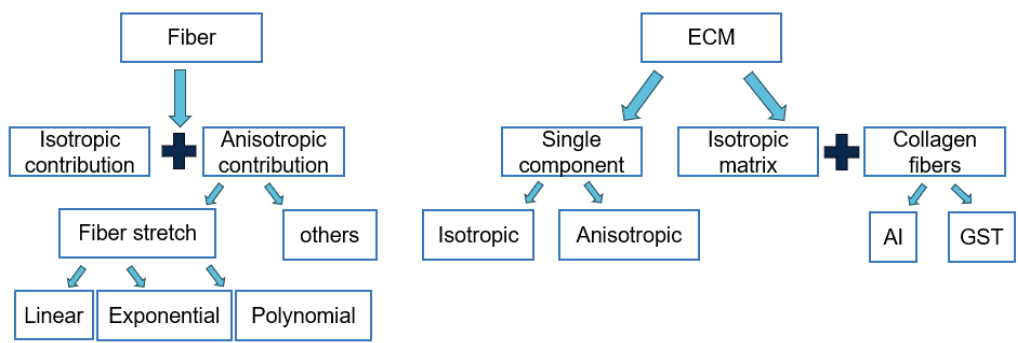

**Fig. 3** Main approaches to represent fibers and ECM behaviors at the microscale, AI: angular integration, GST: generalized structure tensor [descriptive caption: this figure shows a diagram summarizing the different modeling approaches at the microscale. On the first level the distinction is made between muscle fibers and extracellular matrix. On the second level the distinction is made between isotropic and anisotropic contribution for the fibers, whereas the ECM modeling can be distinguished in a single component model, or isotropic matrix plus collagen fibers model. More ramifications give additional precisions.]

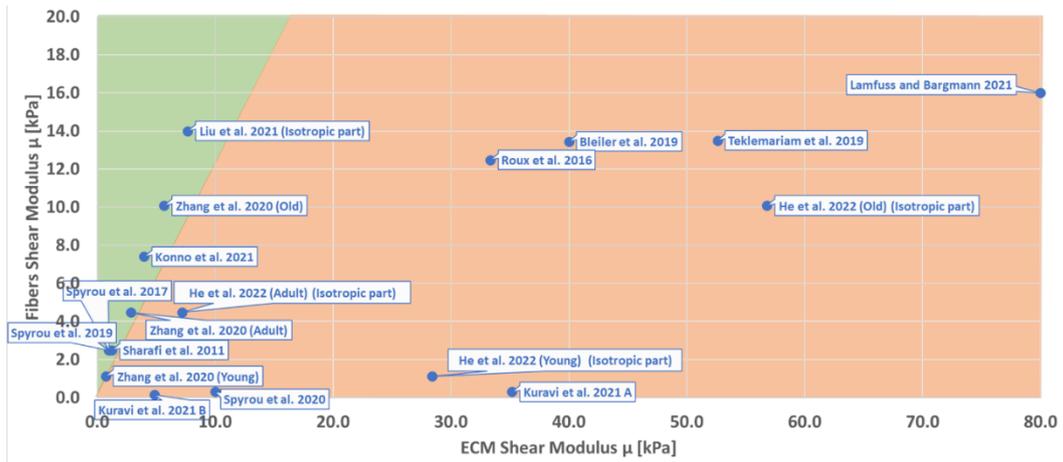

**Fig. 4** Comparison of equivalent shear moduli in small deformations for ECM and muscle fibers in a large number of multiscale models proposed in the literature. The boundary between stiffer fibers (green area) and stiffer ECM (orange area) is illustrated [descriptive caption: this graphic shows scattered points, each representing a different numerical study. The axes are the ECM shear modulus in kPa as the x-axis, and the muscle fiber shear modulus in kPa as the y-axis]

**Table 1** Main experimental studies used to identify the behavioral laws of multiscale numerical models of skeletal muscle (N.I.: no information) [descriptive caption: this table presents characteristics about experimental datasets published in the literature. It includes 5 entries for each dataset, which are the reference of the study, the considered solicitation, the tested species, the tested muscles, and the strain rate used for the experiment]

| Reference | Solicitation | Species | Muscle | Strain rate |
|---|---|---|---|---|
| Hawkins and Bey 1994 [108] | Active and passive response to uniaxial tensile test in the muscle fiber direction | Rat | Tibialis anterior | N.I. |
| Morrow *et al.*, 2010 [106] | Longitudinal extension, transverse extension and longitudinal shear test | Rabbit | Extensor digitorum longus | $0.5\,e^{-3}\,s^{-1}$ |
| Calvo *et al.*, 2010 [105] | *In vitro* and *in vivo* uniaxial tensile tests | Rat | Tibialis anterior | $0.33\,e^{-3}\,s^{-1}$ |
| Meyer and Lieber 2011 [19] | Tensile tests on individual fiber, fiber composite and bundle composite | Mouse | $5^{th}$ toe of extensor digitorum longus | $20\,s^{-1}$ |
| Takaza *et al.*, 2013 [75] | Tensile tests at 0°, 30°, 45°, 60° and 90° to the muscle fiber direction | Pig | Longissimus dorsi | $0.5\,e^{-3}\,s^{-1}$ |
| Mohammadkhah *et al.*, 2016 [109] | Tensile and compression tests at 0°, 45° and 90° to the muscle fiber direction | Chicken | Pectoralis | $0.5\,e^{-2}\,s^{-1}$ |

**Table 2** Appendix A: Main multiscale models [descriptive caption: this table presents characteristics about multiscale models published in the literature in chronological order. It includes 11 entries for each model, which are (1) the reference of the model, (2) a short description of the model, (3) indications about the isotropic part of the model concerning the extracellular matrix, (4) indications about the anisotropic part of the model concerning the extracellular matrix, (5) indications about the isotropic part of the model concerning the muscle fiber, (6) indications about the anisotropic part of the model concerning the muscle fiber, (7) the modeled constituents, (8) the scale transition method, (9) the reference for the experimental datasets used for identification of the model, (10) the FE solver used, and (11) the reference for the experimental datasets used for validation of the model]

| Studies | Description | ECM Law & Parameters | | Fibers Law & Parameters | | Components | Scale transition | Identification | FE solver | Validation |
|---|---|---|---|---|---|---|---|---|---|---|
| | | **Isotropic part** | **Anisotropic part** | **Isotropic part** | **Anisotropic part** | | | | | |
| **Ceelen et al. 2008** [87] | chemomechanical model of hypoxia in muscles due to muscle compression | Neo-Hookean | | Neo-Hookean | | ECM, Fibers, Capillaries | Averaged results | Bosboom et al. (2001), Labbe et al. (1987) | | |
| **Röhrle et al. 2008, 2012** [83,84] | | Passive Mooney-Rivlin (1rst order) | | | Passive & active normalized force length relationships | Fibers, ground matrix, sarcomeres | Concurrent numerical homogenization | Shorten et al. (2007) | Open CMISS | |
| **Sharafi & Blemker 2010** [36] | 3-scale model to study variations of the microstructure | | Model of Blemker et al. (2005) | | Model of Blemker et al. (2005) | Fibers, ECM, fascicles | Periodic homogenization | | | |
| **Sharafi et al. 2011** [92] | Study of force transmission in a realistic fascicle geometry | | Model of Blemker et al. (2005) | | Model of Blemker et al. (2005) | Fibers, ECM | Periodic homogenization | parameters taken from the literature | Matlab, NIKE3D | |
| **Gindre et al. 2013** [88] | tension compression model | | Force generated by wavy collagen helices | | Wormlike chain model of titin | ECM, Fiber, Titin, Collagen | Homogenization: Voigt / mixture law | parameters taken from the literature | Matlab | Takaza et al. (2013), Calvo et al. (2010), Morrow et al. (2010), Nie et al. (2011), Yamada (1970), Martins et al. (1998), Van Loocke et al. (2006), Grieve and Armstrong (1988), Vannah and Childress (1993), Zheng et al. (1999) |
| **Heidlauf et al. 2013** [120] | | Passive Mooney-Rivlin (1rst order) | | Passive Mooney-Rivlin (1rst order) | Passive part model of Markert et al. () + active part model | ECM, Fast & slow fibers, Sarcomeres | Concurrent numerical homogenization | Shorten et al. (2007) | Open CMISS | |

| Reference | Approach | Passive law | Geometry | Passive ECM law | Active law | Constituents | Multiscale approach | Experimental data | Software | Validation |
|---|---|---|---|---|---|---|---|---|---|---|
| | | | | | of Röhrle et al. (2012) | | | | | |
| **Virgilio et al. 2015** [38] | Study of Duchenne Muscular Dystrophy with AI generated geometry | | Model of Blemker et al. (2005) | | Model of Blemker et al. (2005) | Fibers, ECM, fat, dystrophin | Averaged results | | NIKE3D | |
| **Roux et al. 2016** [43] | Discrete Element Method | Hookean accounting for collagen toe region | | Hookean | | ECM, Collagen Fibers, Tendons, Epimysium | Identification | Matscheke et al. 2013 Regev et al. 2011 | GranOO | |
| **Heidlauf et al. 2016, 2017, Schmid et al. 2019** [39–41] | Model based on sliding-filament & cross-bridge theory | Passive Mooney-Rivlin (1rst order) | | Passive Mooney-Rivlin (1rst order) | Model of Heidlauf et al. (2016) + active contribution of actin-titin interaction | ECM, Fast & slow fibers, Sarcomeres Titin | Concurrent numerical homogenization | Shorten et al. (2007) | Open CMISS | |
| **Spyrou et al. 2017** [50] | Homogenized muscle model | Neo-Hookean | Gindre et al. (2013) | Neo-Hookean | Active part: normalized weibull distribution (Ehret 2011) and hyperbolic function Böl et al. (2008), Van Leeuwen et al. (1997) Passive part: linear fonction Blemker et al. (2005) | Fibers, ECM | Analytical homogenization + numerical periodic homogenization | Hawkins 1 Bey 1994, Van Loocke et al. 2006, Böl & Reese 2008, Thacker et al. 2012, Holzapfel et al. 2005, Purslow & Trotter 1994, Lieber et al. 2003 | ABAQUS | Hawkins & Bey 1994 |
| **Marcucci et al. 2017** [34] | Passive 3D FE model based on Hill model | | | | Hill model, using quadratic and exponential formulations for the elements | Fibers | Averaged results | Botinelli et al. 1996, He et al. 2000 | ABAQUS | |
| **Marcucci et al. 2019** [35] | Passive 3D Hill model with separate contributions of muscle fibers and ECM | Isotropic element in parallel to the Hill model, with exponential behavior | | | Hill model, using quadratic and exponential formulations for the elements | Fibers, ECM | Averaged results | Own experiments | ABAQUS | |

| Reference | Description | | | | Additional info | Components | Homogenization | Parameters from | Software | Validation |
|---|---|---|---|---|---|---|---|---|---|---|
| **Bleiler et al. 2019** [76] | Description of the passive behaviour of skeletal muscle tissue | Neo-Hookean | Lanir et al. 1983 | Neo-Hookean | - | Fibers, ECM | Voigt-type homogenization | Holzapfel et Weizsäcker 1998, Holzapfel et al. 2005, Purslow & Trotter 1994, Light & Champion 1984, Light et al. 1985, Mohammadkhah et al. 2018 | Open CMISS | Hawkins & Bey 1994, Morrow et al. 2010, Calvo et al. 2010, Takaza et al. 2013 |
| **Spyrou et al. 2019** [55] | Active skeletal muscle model | Neo-Hookean | Gindre et al. 2013 | Neo-Hookean | Active part: normalized weibull distribution (Ehret 2011) and hyperbolic function Böl et al. (2008), Van Leeuwen et al. (1997) Passive part: linear fonction Blemker et al. (2005) | Fibers, ECM | Analytical homogenization + numerical periodic homogenization | Hawkins & Bey 1994 | ABAQUS | Morrow et al. 2010, Hawkins & Bey 1994 |
| **Teklemariam et al. 2019** [49] | model of motor unit remodeling with age | Mooney-Rivlin (2nd order) | | Yeoh model | Second order polynomial (passive) + active stress in fiber direction | Fibers, ECM, tendon | Averaged results | Bosboom et al. 2001; Chi et al. 2010; Prado et al. 2005 | Comsol Multiphysics | |
| **Carriou et al. 2019** [86] | Hill-inspired model detailing the electromechanical behavior of the muscle based on the Huxley formulation | | | | | Motor Unit contractile elements repeated in parallel to constitute the muscle, including calcium dynamics for activation | Averaged results | Benoussaad et al. 2013 | | Benoussaad et al. 2013 |

| Reference | Focus | Fiber model | ECM model | Matrix model | Active part | Components | Homogenization | Parameters from | Software | Validation |
|---|---|---|---|---|---|---|---|---|---|---|
| **Spyrou et al. 2020** [56] | Muscle fibers, intramuscular connective tissue (IMCT), intramuscular fat (IMF) | Neo-Hookean IMCT | Lanir et al. 1983, Gasser et al. 2006 | Neo-Hookean | standard reinforcement model Lopez-Pamies et al. 2010 | Fibers, ECM, fat | Numerical periodic homogenization | Böl et al. 2019, van der Rijt et al. 2006, Holzapfel et al. 2005, Purslow 1989, Purslow & Trotter 1994, Gefen et Dilmoney 2007 | ABAQUS | Morrow et al. 2010 |
| **Zhang et al. 2020** [37] | Young, Adult & Old models | cubic polynomial hyperelastic model | | Mooney-Rivlin (2nd order) | Model of Blemker et al. 2005 | Fibers, ECM | Averaged results | | | Hawkins & Bey 1994, Morrow et al. 2010 Calvo et al. 2010 |
| **Roux et al. 2021** [44] | Discrete element model of active muscle | Hookean | | Hookean + force length relationship as parabolic curve (Winter et al. 2010, Mohammed et al. 2016) | | Fibers, ECM, tendon | Averaged results | Matscheke et al. 2013 Regev et al. 2011 Winters & Stark 1988 Gordon et al. 1966 Winters et al. 2011 | GranOO | |
| **Lamfuss & Bargmann 2021** [47] | Fascicle modeling, effect of pennation angle | Ogden (1rst order) | | Passive part: Neo Hookean | Active part: Heidlauf & Röhrle 2013 | Fibers, ECM | Numerical homogenization | Haug et al. 2018, Kammoun et al. 2019, Schneidereit et al. 2018 | | Meyer & Lieber 2011 |
| **Liu et al. 2022** [57] | Model to study shear loading | Neo-Hookean | Type I collagen model with exponential + polynomial formulations | Neo-Hookean | Exponential + polynomial formulations | Fibers, ECM | Numerical periodic homogenization | Smith et al. 2011, Böl et al. 2019, Kohn et al. 2021, own experiments | ABAQUS | Morrow et al. 2010, own experiments |
| **Konno et al. 2021** [46] | Model to study cerebral palsy | Yeoh model | Simo et al. 1992 | Yeoh model | Simo et al. 1992 + polynomial passive part + trigonometric active part | ECM, Fibers (Active & passive) | Voigt-type homogenization | Gillies et al. 2011, Winters et al. 2011, Mohammadkhah et al. 2016 | Deal II | Takaza et al. 2013, Mohammadkhah et al. 2016 |
| **Kuravi et al. 2021 a** [32] | Constitutive model of the extracellular matrix | | Rubin & Bodner 2002, Holzapfel et al. 2015 | | Exponential Holzapfel et al. 2000 | ECM, Fibers, Collagen | Numerical periodic homogenization | Böl et al. 2019 | ABAQUS | Van Loocke et al. 2006, Böl et al. 2012, Böl et al. 2019 |

| Reference | Purpose | SEF ECM | SEF Collagen | SEF Fibers passive | SEF Fibers active | Components | Homogenization | Validation | Software | Experimental data |
|---|---|---|---|---|---|---|---|---|---|---|
| **Kuravi et al. 2021 b** [33] | Constitutive model of the extracellular matrix | | Rubin & Bodner 2002, Holzapfel et al. 2015 + contributions for shear modeling | Neo-Hookean | Exponential Holzapfel 2000 | ECM, Fibers, Collagen | Numerical periodic homogenization | Böl et al. 2019, Kohn et al. 2021 | ABAQUS | Own experiments |
| **He et al. 2022** [58] | Young, Adult & Old models | Mooney-Rivlin (3rd order) | Lanir et al. 1983, Bleiler et al. 2019, Spyrou et al. 2019, exponential formulation for collagen helices | Mooney-Rivlin (2nd order) | Exponential passive part + Polynomial active part | ECM, Fibers, Tendons, aponeurosis | Numerical periodic homogenization | Spyrou et al. 2019 Zhang et al. 2020 | ABAQUS | Calvo et al. 2010 |
| **Terzolo et al. 2022** [42] | Model of human vocal folds | Neo-hookean | Orgeas et al. 1998 | Neo-hookean | Orgeas et al. 1998 + repulsive interactions between myofibrils | Fibers, ECM, myofibrils, collagen | Caillerie et al. 2006 | Cochereau et al. 2020 | | |
| **Lamfuss et al. 2022** [48] | Model of damage initiation & rupture of activated muscle fibers resulting from eccentric contractions formation of microcracks | Endomysium Ogden (1rst order) | | Passive part: Neo Hookean | Active part: Heidlauf & Röhrle 2013 + titin force enhancement (Cankaya et al. 2021) + anisotropic damage model | Fibers, ECM | Numerical homogenization | Haug et al. 2018, Meyer et Lieber 2011; Meyer et al. 2011, Schneidereit et al. 2018, Cankaya et al. 2021 | | |
| **Lamfuss et al. 2023** [60] | Model of thermal treatments | Ogden (1rst order) Thermomechanical & thermal energies | | Neo Hookean with thermomechanical & thermal parts | Force length relationship (zuurbier 1995) Thermomechanical & thermal energies, endothermal contribution due to cross-bridge cycling | Fibers, ECM | Numerical homogenization | Erdman & Gos 1990, Okamoto & Saeki 1964, Werner & Buse 1988, Briese 1998, Mcintosch & Anserson 2010, Mutungi & Ranatunga 1998, Segal et al. 1986, Ranatunga 1994, Bershitsky & Tsaturyan 1989, Ried et al. 1999 | | |
| **Sahani et al. 2024** [59] | Model of Duchenne Muscular Dystrophy in mice | - | Angular Integration model of Ateshian et al. 2009, power law based on ellipsoidal collagen fiber distribution | Model of Blemker et al. 2005 | Model of Blemker et al. 2005 | ECM (endomysium, epimysium), Fibers | Periodic homogenization then Voigt homogenization to include epimysium | Sahani et al. 2022, Wohlgemuth et al. 2023, Morrow et al. 2010, Zajac et al. 1989, own experiments | FeBio | own experiments |